\documentclass[aps,pra,twocolumn,superscriptaddress]{revtex4-2}

\usepackage{graphicx}
\usepackage{amsmath}
\usepackage{textcomp}
\usepackage{pifont}
\usepackage{tikz}
\usepackage{blindtext}
\newcommand{\includelabeledgraphics}[4][]{%
\begin{tikzpicture}%
\node (1) [inner sep=0, outer sep=0] {\includegraphics[#1]{#2}};%
\node at (1.#3) [anchor=#3,fill=white,draw=black] {#4};%
\end{tikzpicture}%
}
\usepackage[colorlinks]{hyperref}

\begin{document}

 \title{Site-specific plan-view TEM lamella preparation of pristine surfaces with a large field of view}

 \author{Tobias Meyer}
 \author{Tobias Westphal}
 \affiliation{4th Institute of Physics -- Solids and Nanostructures, University of G\"ottingen, Friedrich-Hund-Platz 1, 37077 G\"ottingen, Germany}

 \author{Birte Kressdorf}
 \author{Ulrich Ross}
 \author{Christian Jooss}
 \affiliation{Institute of Materials Physics, University of G\"ottingen, Friedrich-Hund-Platz 1, 37077 G\"ottingen, Germany}

 \author{Michael Seibt}
 \email{mseibt@gwdg.de}
 \affiliation{4th Institute of Physics -- Solids and Nanostructures, University of G\"ottingen, Friedrich-Hund-Platz 1, 37077 G\"ottingen, Germany}


 \begin{abstract}
 Transmission electron microscopy has become a major characterisation tool with an ever increasing variety of methods being applied in wide range of scientific fields. However, the probably most famous pitfall in related workflows is the preparation of high-quality electron-transparent lamellae enabling for extraction of valuable and reliable information.
   Particularly in the field of solid state physics and materials science, it is often required to study the surface of a macroscopic specimen with plan-view orientation.
   Nevertheless, despite tremendous advances in instrumentation, i.e. focused ion beam, the yield of existing plan-view lamellae preparation techniques is relatively low compared to cross-sectional extraction methods. Furthermore, techniques relying on mechanical treatments, i.e. conventional preparation, compromise site-specifity. In this paper, we demonstrate that by combining a mechanical grinding step prior to backside lift-out in the focused ion beam plan-view lamellae preparation becomes increasingly easy. The suggested strategy combines site-specifity with micrometer precision as well as possible investigation of pristine surfaces with a field of view of several hundred square micrometers.
 \end{abstract}

 \maketitle

\section{Introduction}

While the preparation of high-quality site-specific cross-sectional transmission electron microscopy (TEM)
lamellae has become rather straight forward due to the developments in focused ion beam (FIB) instrumentation,
the extraction of plan-view lamellae matching the same criteria remains comparatively difficult.
Consequently, many different approaches employing mechanical and FIB treatments or a combination of both
to achieve high-quality plan-view TEM foils have been reported. To the best of our knowledge, these 
approaches can 
generally be categorized into two groups: (i) A mechanical treatment - e.g. embedding into
resin~\cite{Lemrik2014}, sandwich gluing
~\cite{Stevie2008}, tripod 
respectively planar polishing~\cite{Anderson1997,OShea2014} - optionally followed by FIB slicing at the edge of the residual 
bulky sample. (ii) A direct extraction of a block of material from the top using the FIB~\cite{Langford2001}.
Whereas category (i) clearly compromises site-specifity since lamellae can only be extracted at
a mechanically prepared edge, the main drawback of category (ii) is the rather complicated and time-consuming excavation as well as the necessity to protect the surface
both from ion beam irradiation as well as contamination due to redeposition. Possible solutions are the 
use of a protective block which is mounted with a micromanipulator~\cite{Jublot2014} or a shielding platinum 
wall~\cite{Li2017}. While in the latter case, residual surface contaminants ultimately had to be removed by FIB milling 
leading to an altered surface, the authors of~\cite{Jublot2014} managed to remove exclusively the protective block
during the thinning process 
revealing the pristine surface.

In this paper, we demonstrate a novel strategy of combining mechanical polishing and FIB lift-out from the backside
to prepare plan-view TEM lamellae of pristine surfaces as well.
Compared to those in category (i), the presented method offers uncompromised site-specifity. 
In addition, the approach avoids the experimental difficulties of cutting a bulky block from the 
sample's top as it is done in category (ii) methods enabling for TEM lamellae preparation with a large field 
of view. 


\section{Instrumentation}

Lamellae extraction and thinning has been performed in an FEI Helios G4 Dual Beam FIB.
TEM investigations have been conducted in an image-corrected FEI Titan 80-300 operated at 300\,kV and equipped with a 
Gatan Quantum 965 ER image filter for electron energy loss spectroscopy (EELS) data acquisition.
During the scanning TEM (STEM) investigations with an approximate probe size of 1.3\,\r{A} according to the instrument's specifications, the beam current was tuned to 42\,pA for imaging and 150\,pA 
for EELS mapping. The acceptance semi-angle of the spectrometer was set to 39\,mrad and the inner and outer
collection semi-angle of the included annular dark-field (ADF) detector to 46.8\,mrad and 200\,mrad.

\section{Pristine homogeneous thin film preparation}

\begin{figure}
\centering
\includelabeledgraphics[width=.46\textwidth]{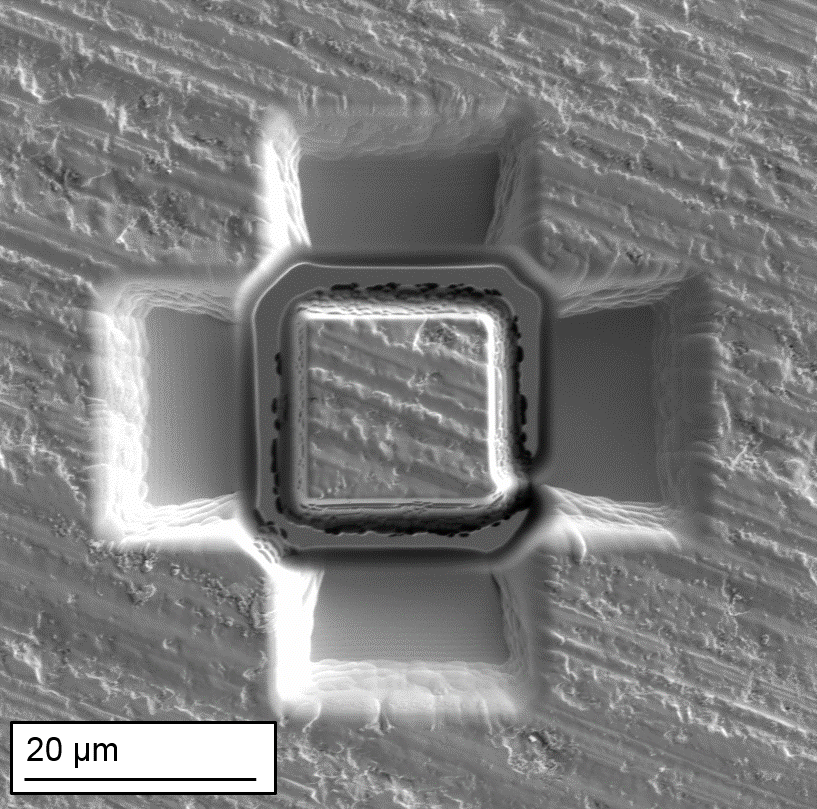}{north west}{(a)}
\includelabeledgraphics[width=.46\textwidth]{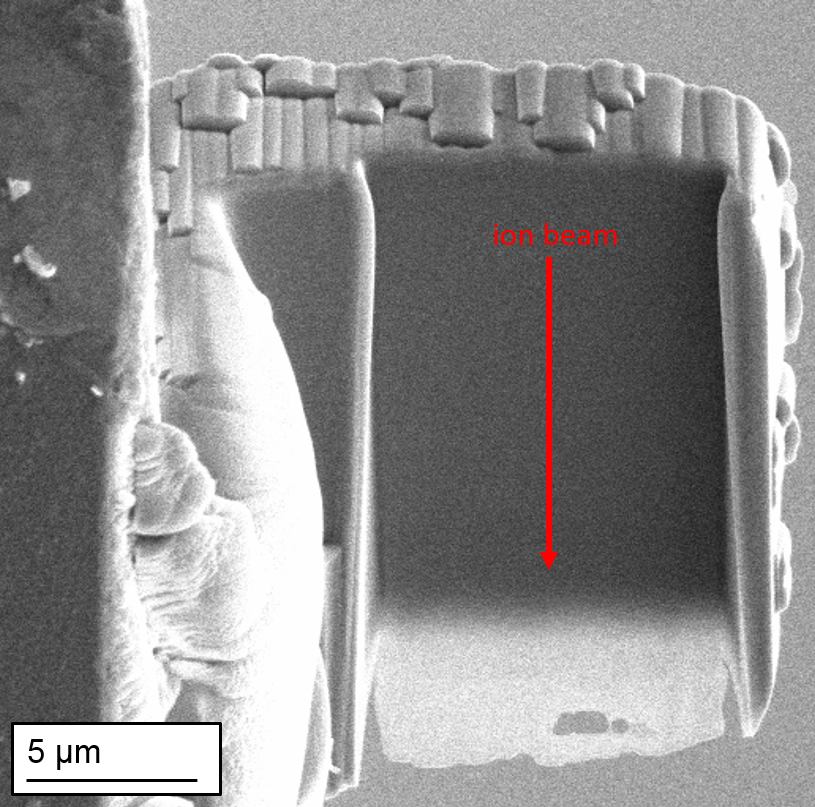}{north west}{(b)}
\caption{(a) Trenches milled in the FIB for backside lift-out using regular cross-sections (outer regions) and rectangles (inner cutting frame). A small bridge is left until the nanomanipulator is attached for transfer. (b) Final lamella with electron transparent region at the bottom. During thinning, the ion beam marked with the red arrow impinges with an offset angle of 15$^\circ$ with respect to the pristine surface resulting in a thicker part at the top while milling the bottom part to electron transparency. All relevant stage positions during the FIB preparation are summarized in the supplementary material.}
\label{fig_sem}
\end{figure}

A nominally 100\,nm thick Pr$_{1-x}$Ca$_x$MnO$_3$ (PCMO) film with $x=0.34$ has been deposited
via ion beam sputtering on a SrTi$_{1-y}$Nb$_y$O$_3$ (STNO) substrate with $y=0.002$ as it has been described
in~\cite{ifland2017contribution}.
Subsequently, the sample was mounted upside down on a stainless steel cylinder using QuickStick 135 Mounting Wax
and ground to a thickness of approximately 20\,\textmu m with a PRESI Minitech 233 disc grinder and 
P2400 grinding paper.
The wax was removed in an acetone bath and the specimen was glued upside down with silver paint  
on a scanning electron microscope (SEM) stub such that the region of interest sticks
over the stub's edge. 

After insertion into the FIB, four regular cross sections (following the FEI terminology) with dimensions
$20\times14\times50$\,\textmu m$^3$ and a beam current of 47\,nA were dug from the backside to create the outer trenches with fourfold symmetry
in Fig. \ref{fig_sem} (a). Please note that the indicated depth of the regular cross section is much larger
than the sample thickness because the value is calibrated to silicon with a significantly higher sputtering rate
compared to STNO. 
In a second step, the inner cut-out frame in Fig. \ref{fig_sem} (a) was created employing rectangular patterns 
at 47\,nA. In the lower right part, a small bridge was left until the central block was attached to the 
EasyLift\textsuperscript{TM} nanomanipulator. Finally, after lifting out and attaching 
the block to an Omniprobe support grid with platinum, a 15\,\textmu m wide and 50\,\textmu m deep cleaning cross 
section was first used with 21\,nA and subsequently with 2.5\,nA to thin the block to electron transparency as it is shown in Fig. \ref{fig_sem} (b). The red arrow indicates the ion milling direction. Importantly, the sample was intentionally tilted such that the angle between the ion beam and the surface amounts to 15$^\circ$ to maintain a rather thick part at the top guaranteeing mechanical stability during the substrate removal.  

\begin{figure}
\includelabeledgraphics[width=.45\textwidth]{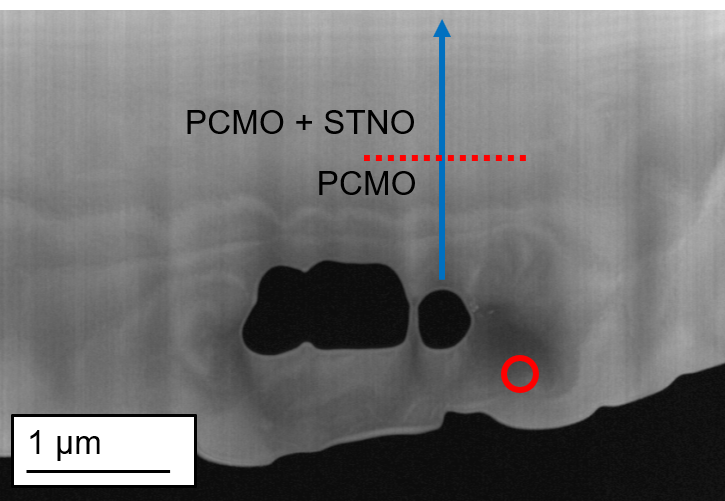}{north west}{(a)}
\includelabeledgraphics[width=.46\textwidth]{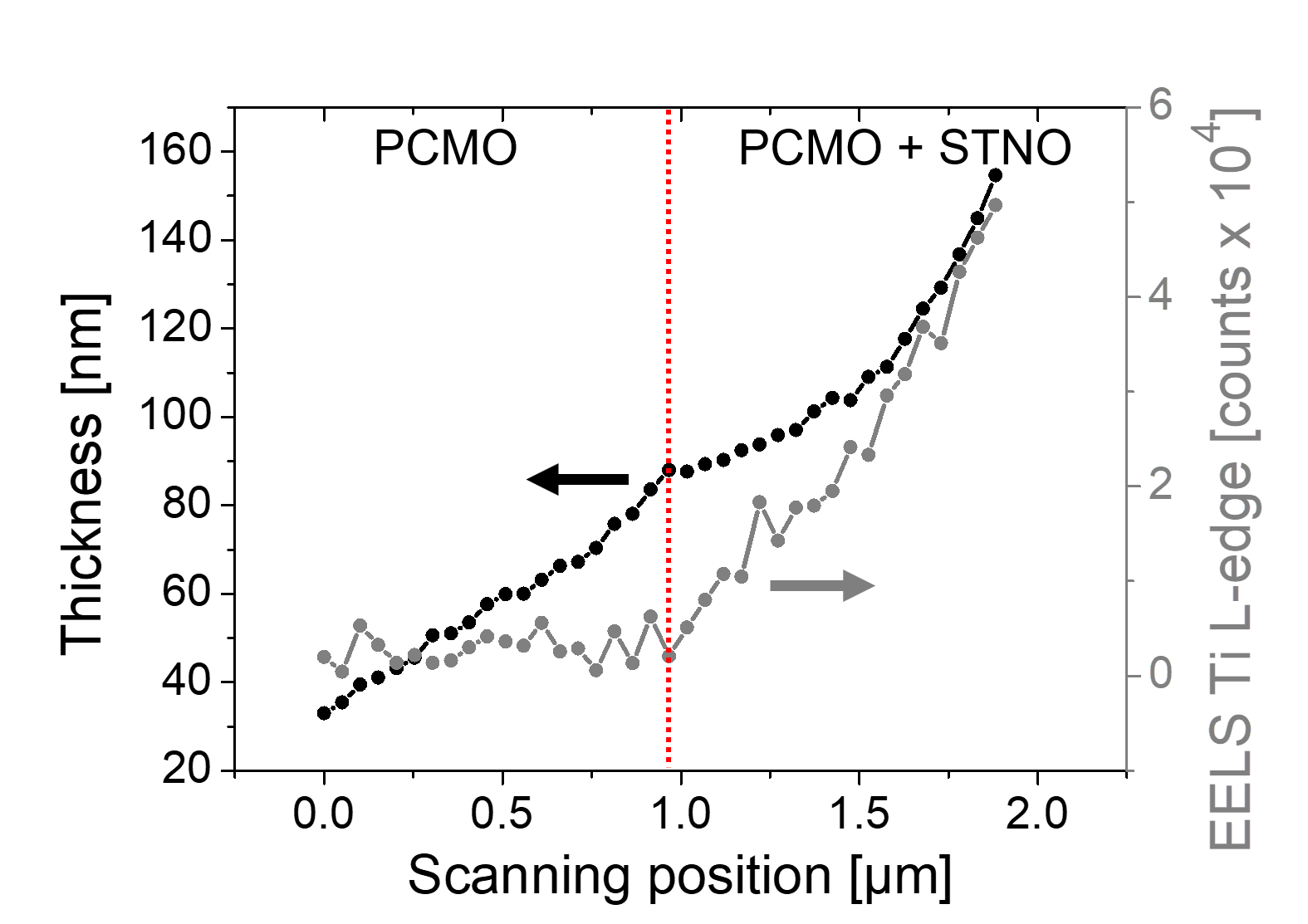}{north west}{(b)}
\caption{(a) ADF-STEM overview of the thin region in Fig. \ref{fig_sem} (b). The blue arrow indicates the profile along which the EELS data shown in (b) have been recorded. Furthermore, the red circle marks the region investigated by SAED and HRSTEM. (b) Thickness as well as integrated Ti L-edge intensity after power-law background subtraction along the blue arrow in (a). The red dotted lines mark the position at which substrate contributions become significant.}
\label{eels}
\end{figure}
\begin{figure}[h]
\includelabeledgraphics[width=.46\textwidth]{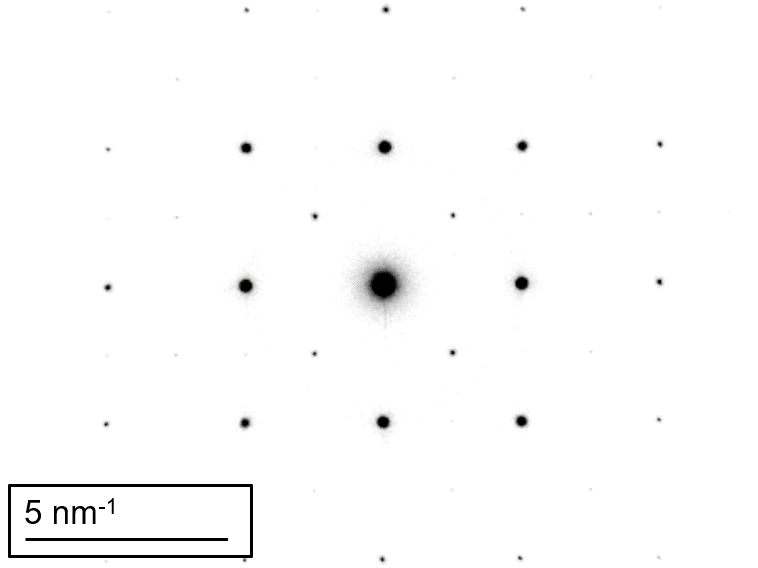}{north west}{(a)}
\includelabeledgraphics[width=.46\textwidth]{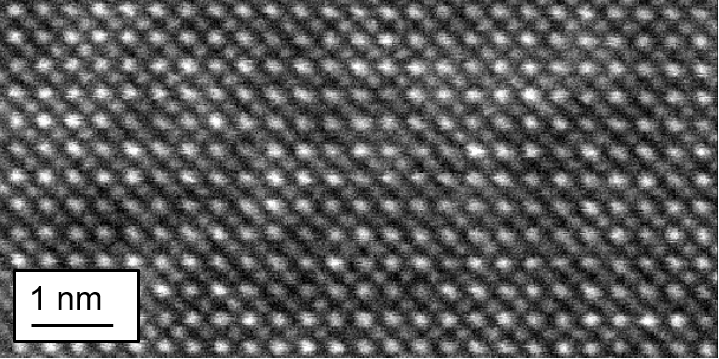}{north west}{(b)}
\caption{SAED pattern (a) as well as ADF-STEM acquisition (b) of the region marked with the red circle in Fig. \ref{fig_sem} showing high crystalline quality and no significant signal from contamination or disordered layers.}
\label{diff}
\end{figure}

The resulting lamella was investigated in the TEM by means of ADF-STEM combined with EELS as well as selective area electron diffraction (SAED) to assess its quality. Fig. \ref{eels} (a) shows an ADF-STEM overview of the thin part of the specimen including a red arrow resp. circle marking the areas analyzed with EELS resp. SAED as well as high-resolution STEM (HRSTEM). 
The results of the former are presented in Fig. \ref{eels} (b) showing both the corresponding thickness as well as the integrated counts of the Ti L-edge after power-law background subtraction. Please note that $\lambda_\text{PCMO}=110$\,nm~\cite{Norpoth} was used to obtain absolute thicknesses following Malis' method~\cite{Malis} yielding incorrect values in regions with remaining STNO subtrate. The point along the line profile where substrate contributions get significant is marked with a dotted red line. In fact, the kink in the thickness curve caused by the change in lambda (and possibly the thinning behaviour) is both consistent with the emergence of finite Ti L-edge counts as well as the nominal film thickness. Assuming that the kink position follows roughly the horizontal direction across the 15\,\textmu m wide field of view (which can be confirmed by a two-dimensional thickness map) leads to an estimated area of 20\,\textmu m$^2$ of free-standing PCMO film. 
In order to confirm that the sample is sufficiently clean for TEM investigation, i.e. no significant redeposited material or residual wax is present at the surface, an SAED pattern as well as an HRSTEM image of the thin area ($t\approx20$\,nm) marked with the red circle in Fig. \ref{eels} (a) are shown in Fig. \ref{diff} (a) resp. (b).
Even though the contrast in (a) is digitally enhanced as can be inferred from the spurious intensity around the direct beam, no significant contribution of amorphous or poly-crystalline contaminants is observed. Consequently, the prepared free-standing film is well-suitable for HRSTEM imaging as demonstrated in Fig. \ref{diff} (b). Furthermore, it is worth mentioning that in case materials exhibiting redeposition exist, the presented method could be easily combined with soft surface protection layers which can be removed by plasma-cleaning subsequently as demonstrated on cross-sectional lamellae in~\cite{roddatis2019}.

\section{Site-specific preparation with large field of view}

\begin{figure}
\includelabeledgraphics[width=.45\textwidth]{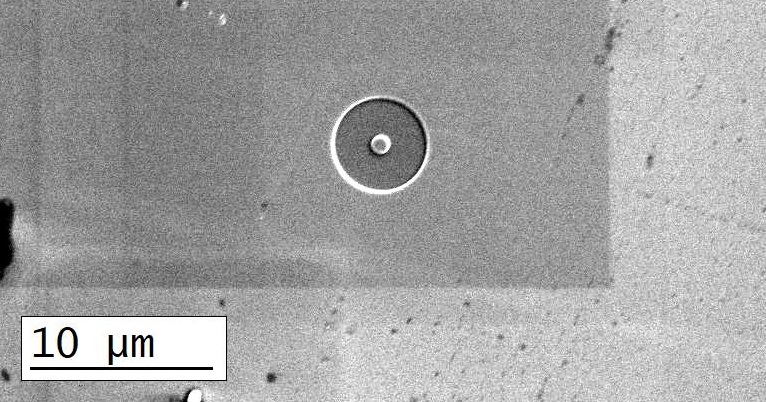}{north west}{(a)}
\includelabeledgraphics[width=.45\textwidth]{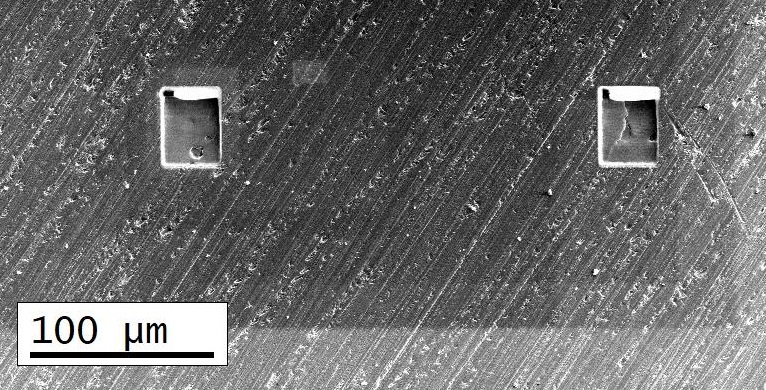}{north west}{(b)}\\
\includelabeledgraphics[width=.45\textwidth]{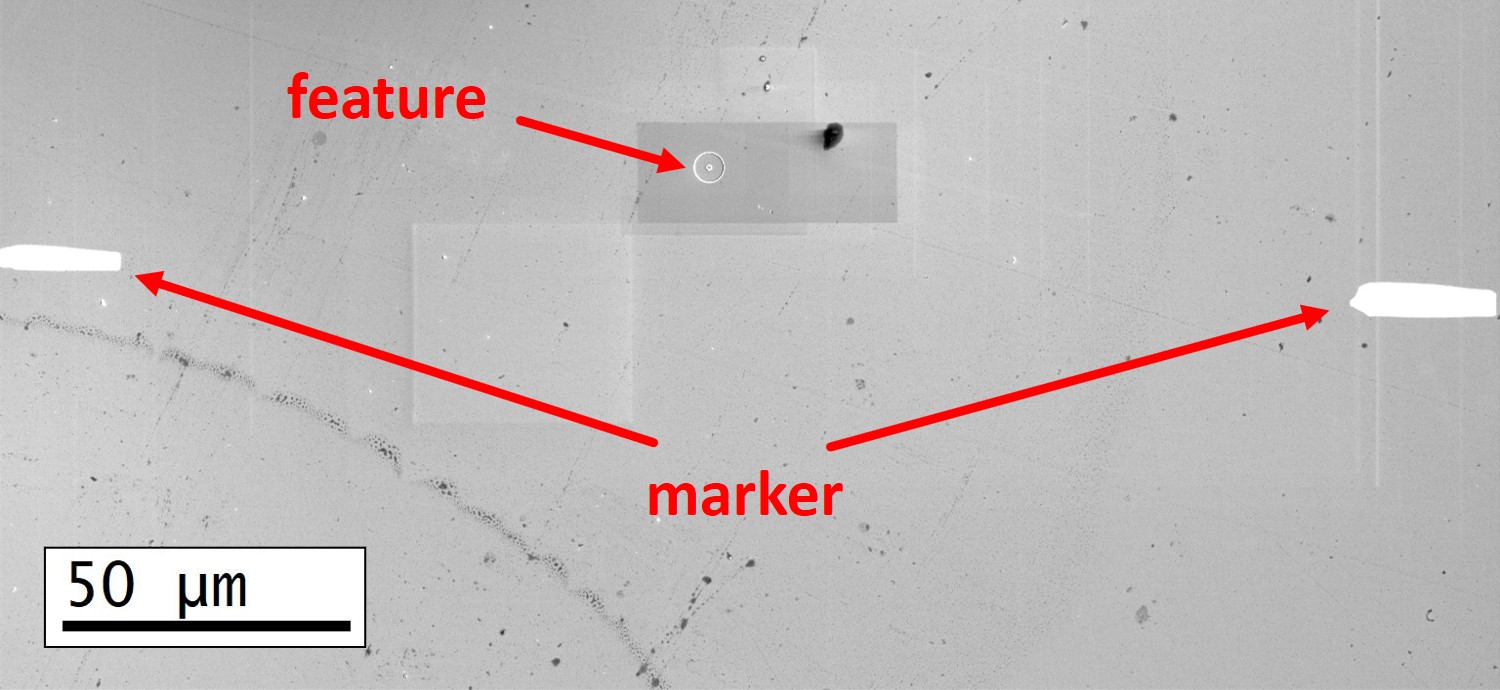}{north west}{(c)}
\includelabeledgraphics[width=.45\textwidth]{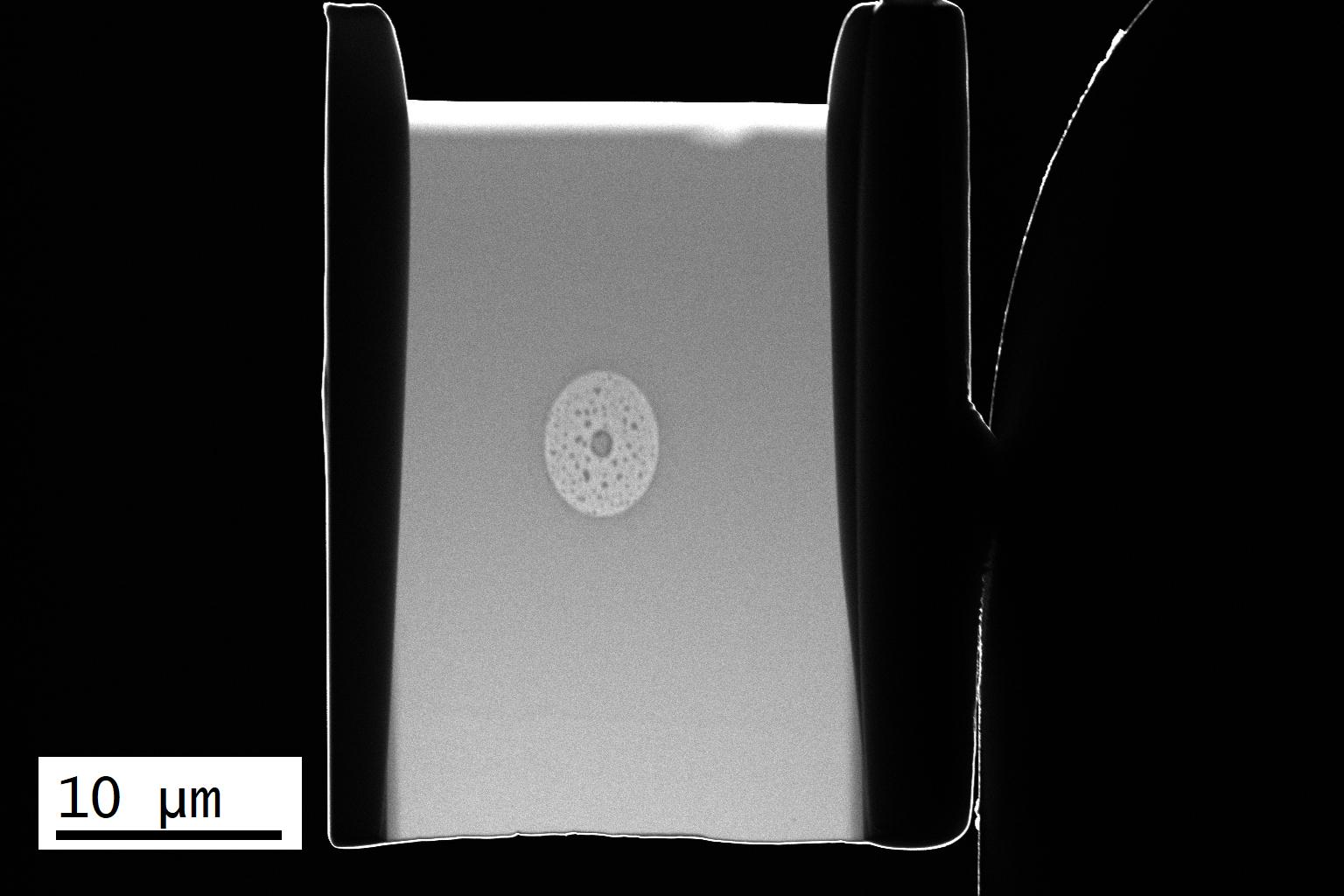}{north west}{(d)}
\caption{(a) Distinct annular feature milled in the silicon surface using the FIB. (b) Regular cross-section markers milled from the backside enabling for (c) localisation of the feature's position with respect to the markers after flipping the sample. (d) Final lamella after backside lift-out and thinning imaged with the STEM detector of the SEM at 30\,kV.}
\label{sitespec}
\end{figure}

In order to prove that the suggested method is not only suitable for extraction of homogeneous materials but can be conducted in a site-specific manner as well, a second example involving a distinct feature at the surface shall be presented. Furthermore, its applicability to another material, i.e. silicon, as well as the scalability to large fields of view will be demonstrated.

Analogously to the mechanical preparation above, a silicon waver has been ground to a thickness of approximately 20\,\textmu m. Subsequently, the specimen was glued across a $2\times1$\,mm$^2$ copper slot enabling the observations in the FIB from the front- and backside. To create a distinct feature at the surface, a 150\,nm deep annular pattern with an inner resp. outer diameter of 1 resp. 5\,\textmu m was dug with the ion beam as shown in Fig. \ref{sitespec} (a).
Subsequently, the following three-step procedure was conducted to extract the unique feature: Firstly, two coarsely-positioned markers (regular cross-sections) were milled with the ion beam from the backside (Fig. \ref{sitespec} (b)). Secondly, the sample was flipped and solely observed with the electron beam to locate the position of the surface feature with respect to the markers (Fig. \ref{sitespec} (c)). Lastly, after flipping the specimen again, the site of interest was extracted by exploiting the information obtained in the previous step. Here, in order to prove the comparatively easy preparation of large lamellae, a block with a $30\times30$\,\textmu m$^2$ surface area was transferred and the angle between ion beam and surface during thinning was decreased to 3$^\circ$.
Fig. \ref{sitespec} (d) shows a high-angle annular dark-field (HAADF) image of the final lamella recorded with the STEM detector included in the SEM at an acceleration voltage of 30\,kV. Obviously, the donut-shaped site of interest is well-located in the extracted block. In fact, comparing its central position to the geometrical mean of the four corners of the approximately $600$\,\textmu m$^2$ large electron transparent area results in a distance of only $1.3$\,\textmu m.

\section{Summary and Conclusion}

In this paper, we have demonstrated that the combination of mechanical grinding and backside lift-out in the FIB is well-suitable for plan-view lamellae extraction of pristine and clean surfaces and meets many requirements of TEM sample preparation simultaneously:
Sticking to the categorisation mentioned above, the suggested procedure clearly circumvents the site-specifity restrictions of group (i) methods as has been demonstrated by the extraction of a distinct surface feature with micrometer precision.
In addition, it enables straight-forward lamellae fabrication with a large field of view since the sputtering volume scales simply with the circumference of the extracted block, i.e. with the square of the edge for quadratic fields of view. In contrast, using category (ii) methods and thus a frontside lift-out, the milling depth in the FIB scales with the edge length resulting in a sputtering volume proportional to the cube of the edge. Consequently, the presented approach combines the benefits of both groups adding value to existing strategies of plan-view TEM lamellae preparation.

\section{Acknowledgments}

The project was funded by the Deutsche Forschungsgemeinschaft (DFG, German Research Foundation) project no. SE 560-6/1 as well as 217133147/SFB 1073, projects B02, Z02.

\section{Author Contributions}

The original idea was conceived by T.M.; Polishing, FIB preparation, and TEM investigation of the samples was conducted by T.M., T.W., and U.R.; The thin film perovskite sample was prepared by B.K.; The manuscript was written by T.M. and T.W. under revision of C.J. and M.S.; All authors read and agreed on the written paper and declare no conflicts of interest;


\end{document}